\documentclass[aps,prd,twocolumn,nofootinbib,superscriptaddress]{revtex4}
\usepackage{txfonts}
\usepackage{graphicx}
\usepackage{dcolumn}
\usepackage{bm}
\usepackage{amssymb}
\usepackage{latexsym}
\usepackage[colorlinks, linkcolor=blue, citecolor=blue, urlcolor=blue]{hyperref}

\begin{document}

\title{Holographic Ricci dark energy in Randall-Sundrum braneworld:
Avoidance of big rip and steady state future}

\author{Chao-Jun Feng}
\email{fengcj@itp.ac.cn} \affiliation{Shanghai United Center for
Astrophysics (SUCA), Shanghai Normal University, %100 Guilin Road,
Shanghai 200234, China} \affiliation{Institute of Theoretical
Physics, Chinese Academy of Sciences, Beijing 100190, China}

\author{Xin Zhang}
\email{zhangxin@mail.neu.edu.cn} \affiliation{Department of Physics,
College of Sciences, Northeastern University, Shenyang 110004,
China} \affiliation{Kavli Institute for Theoretical Physics China
(KITPC), Chinese Academy of Sciences, PO Box 2735, Beijing 100190,
China}

%\date{\today}

\begin{abstract}
In the holographic Ricci dark energy (RDE) model, the parameter
$\alpha$ plays an important role in determining the evolutionary
behavior of the dark energy. When $\alpha<1/2$, the RDE will exhibit
a quintom feature, i.e., the equation of state of dark energy will
evolve across the cosmological constant boundary $w=-1$.
Observations show that the parameter $\alpha$ is indeed smaller than
$1/2$, so the late-time evolution of RDE will be really like a
phantom energy. Therefore, it seems that the big rip is inevitable
in this model. On the other hand, the big rip is actually
inconsistent with the theoretical framework of the holographic model
of dark energy. To avoid the big rip, we appeal to the extra
dimension physics. In this Letter, we investigate the cosmological
evolution of the RDE in the braneworld cosmology. It is of interest
to find that for the far future evolution of RDE in a
Randall-Sundrum braneworld, there is an attractor solution where the
steady state (de Sitter) finale occurs, in stead of the big rip.

\end{abstract}


\maketitle

Many experiments like the observations of Type Ia supernovae (SN)
\cite{Riess:1998cb}, cosmic microwave background (CMB)
\cite{Spergel:2006hy} and large scale structure (LSS)
\cite{:2007wu}, etc. have strongly confirmed that our universe is
undergoing an accelerated expansion. However, since ordinary matters
such as stars always attract each other due to the Newton's gravity,
it was expected that the universe can only experience a decelerated
expansion. Thus, there must be an unknown energy component with
exotic properties, such as repulsive gravity, living in the
universe, and people often call it dark energy. Experiments have
indicated that today there are about $73\%$ dark energy and $27\%$
matter components in the universe. However, so far, people still do
not understand what dark energy is, from the fundamental theory (for
reviews see, e.g., \cite{DErev}).

It seems that the preferred candidate for dark energy is the
Einstein's cosmological constant, but it is very difficult to
understand in the modern field theory; for example, it suffers from
the fine-tuning and the coincidence problems. The fine-tuning
problem asks why the vacuum energy density observed today is so much
smaller than the value predicted by the quantum field theory, and we
thus need to fine tune the cosmological constant to cancel it and
get the observational value; the coincidence problem asks why it is
just today that dark energy becomes important, though its
evolutionary behavior is so different from that of ordinary matter.
In order to eliminate these problems, a lot of dynamical dark energy
models have been built, such as quintessence, phantom, and quintom
models that are basically scalar field models. On the other hand,
there is another way of explaining the accelerated expansion of the
universe. In this way, the cosmic acceleration is explained by some
geometric effects; for example, perhaps the Einstein's gravitational
theory should be modified, such as the $f(R)$ theory and the DGP
model, etc..

It is well known that the cosmological constant (or the vacuum
energy density) is actually closely related to an ultraviolet (UV)
problem in the quantum field theory. A simple evaluation in quantum
field theory leads to a discrepancy of 120 orders of magnitude,
comparing the theoretical result with the observation. Obviously,
the key point is the gravity. In a real universe, the effects of
gravity should be involved in this evaluation. So, actually, the
cosmological constant (or dark energy) problem is in essence an
issue of quantum gravity \cite{Witten:2000zk}. However, by far, we
have no a complete theory of quantum gravity, so it seems that we
have to consider the effects of gravity in some effective field
theory in which some fundamental principles of quantum gravity
should be taken into account. It is commonly believed that the
holographic principle \cite{holop} is just a fundamental principle
of quantum gravity.

The holographic principle is expected to play an important role in
dark energy research. When considering gravity, namely, in a quantum
gravity system, the conventional local quantum field theory will
break down due to the too many degrees of freedom that would cause
the formation of black hole. So, there is a proposal
\cite{CohenZX:1999} saying that the holographic principle may put an
energy bound on the vacuum energy density, $\rho_{vac} L^3 \leq
LM^2_{pl}$, where $\rho_{vac}$ is the vacuum energy density and
$M_{pl}$ is the reduced Planck mass. This bound says that the total
energy in a spatial region with size $L$ should not exceed the mass
of a black hole with the same size. Evidently, this bound implies a
UV/IR duality. Therefore, the holographic principle may lead to a
dark energy model that is actually based on the effective field
theory with a UV/IR duality. From this UV/IR relation, the UV
problem of dark energy can be converted into an IR problem.

By introducing a dimensionless parameter $c$, one can saturate that
bound and write the dark energy density as
$\rho_{de}=3c^2M_{pl}L^{-2}$. The parameter $c$ is introduced to
characterize all of the uncertainties of the theory. Now, the
problem becomes how to choose an appropriate IR cutoff for the
theory. A natural choice is the Hubble length of the universe,
however, it has been proven that there is no cosmic acceleration for
this choice \cite{Hsu04}. Li proposed that \cite{Li:2004rb}, instead
of Hubble horizon, one can choose the event horizon of the universe
as the IR cutoff of the theory. This choice not only gives a
reasonable value for dark energy density, but also gives rise to an
acceleration solution for the cosmic expansion.

Although the holographic model based on the future event horizon is
successful in fitting the current data, some authors asked why the
current acceleration of the universe is determined by its future.
Actually, the future event horizon as the IR cutoff is not the
unique choice for the holographic dark energy model. There are other
versions of holographic dark energy, see, e.g., \cite{agegraphic}.
Also motivated by the holographic principle, Gao et al.
\cite{Gao:2007ep} proposed the holographic Ricci dark energy (RDE)
model recently, in which the future event horizon area is replaced
by the inverse of Ricci scalar, and this model is also
phenomenologically viable.

Assuming the maximal black hole in the universe is formed by
gravitation collapsing of the perturbations in the universe, then
the ``Jeans'' scale of the perturbations gives a causal connection
scale $R_{CC}$ that is naturally to be chosen as an IR cutoff in the
holographic setup \cite{Cai:2008nk}. For tensor perturbations, i.e.,
gravitational perturbations, $R_{CC}^{-2} = Max(\dot H + 2H^2, -\dot
H)$ for a flat universe, and according to Ref.~\cite{Cai:2008nk},
only in the case of $R_{CC}^{-2} =\dot H + 2H^2$, it could be
consistent with the current cosmological observations when the
vacuum density appears as an independently conserved energy
component. Therefore, if one chooses the casual connection scale
$R_{CC}$ as the IR cutoff in the holographic theory of dark energy,
the holographic Ricci dark energy model is then obtained. For recent
progress on the models of holographic dark energy, see, e.g.,
Refs.~\cite{holoext,ageext,ricciext,Zhang:2009un}.

In the Friedmann-Robertson-Walker (FRW) universe, the Ricci scalar
is $\mathcal {R}=-6(\dot H + 2H^2+ k/a^2)$, and the energy density
of RDE reads
\begin{equation}\label{energy density}
\rho_{de} = 3\alpha \left(\dot H + 2H^2 + \frac{k}{a^2}\right)
\propto \mathcal{R}\,,
\end{equation}
where $H = \dot a / a$ is the Hubble parameter, and we have set
$8\pi G = 1$. Here $\alpha=c^2$ is a dimensionless parameter which
plays a significant role in determining the evolutionary behavior of
the RDE. Now, the Friedmann equation reads
\begin{equation}\label{Friedmann}
r^2 = \Omega_{m0}e^{-3x} +
(1-\alpha)\Omega_{k0}e^{-2x}+\alpha\left(\frac{1}{2}\frac{dr^2}{dx}
+ 2r^2\right) \,,
\end{equation}
where $r\equiv H/H_0$, $\Omega_{m0} = \rho_{m0}/(3H_0^2)$,
$\Omega_{k0} = -k/H_0^2$ and $x = \ln a$. Here and after, the
subscript `0' denotes the present values of variables. Solving this
equation, one obtains
\begin{equation}\label{Solution}
r(a)^2 = \Omega_{m0}a^{-3} + \Omega_{k0}a^{-2} +
\frac{\alpha}{2-\alpha}\Omega_{m0}a^{-3} +
f_0a^{-\left(4-\frac{2}{\alpha}\right)},
\end{equation}
where $f_0$ is an integration constant that can be determined, by
the initial condition, as
\begin{equation}
f_0 = 1-\Omega_{k0} - \frac{2}{2-\alpha}\Omega_{m0} \,.
\end{equation}

From Eq.~(\ref{Solution}), one can easily reads the contribution of
RDE, namely, the last two terms on the rhs. We now define
\begin{equation}
\tilde{\Omega}_{de} \equiv \frac{\rho_{de}}{3H_0^2} =
\frac{\alpha}{2-\alpha}\Omega_{m0}a^{-3} +
f_0a^{-\left(4-\frac{2}{\alpha}\right)} \,,
\end{equation}
then from the energy conservation law, we can get the equation of
state,
\begin{equation}
w(z) = -1 + \frac{(1+z)}{3}\frac{d\ln\tilde{\Omega}_{de}}{dz}.
\end{equation}
Since the current observations strongly favor a spatially flat
universe that is also supported by the inflation theory, hereafter
the discussions will be restricted to the case of $\Omega_{k0}=0$
(or $k=0$).

\begin{figure}[h]
\includegraphics[width=0.4\textwidth]{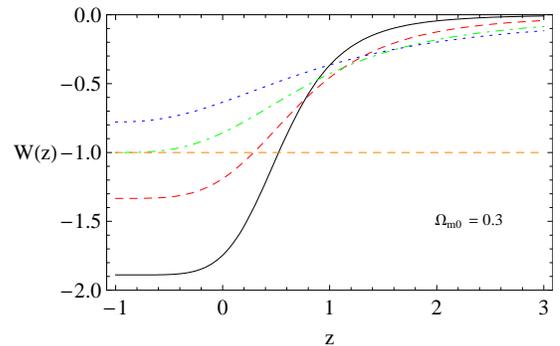}
\caption{\label{fig::eos} Equation of state parameter $w$ for RDE
with parameter $\alpha = 0.3$ (solid curve), $\alpha = 0.4$ (dashed
curve), $\alpha = 0.5$ (dotdashed curve), and $\alpha = 0.6$ (dotted
curve), respectively. The horizontal dashed line corresponds to the
cosmological constant, $w=-1$. Here, we have neglected the spatial
curvature. }
\end{figure}

We have plotted the equation of state parameter $w$ with respect to
the redshift $z$ in Fig.~\ref{fig::eos}. Obviously, the value of
parameter $\alpha$ is very important in determining the evolutionary
behavior of RDE. When $1/2\leq\alpha\leq1$, the equation of state of
dark energy will evolve in the region of $-1\leq w\leq -1/3$. In
particular, if $\alpha=1/2$ is chosen, the behavior of the RDE will
be more and more like a cosmological constant with the expansion of
the universe, such that ultimately the universe will enter the de
Sitter phase in the far future. When $\alpha<1/2$, the RDE will
exhibit a quintom-like evolution behavior (for ``quintom'' dark
energy, see, e.g., \cite{quintom}),\footnote{For early observational
evidence for this kind of dark energy, see also \cite{Alam:2004jy}.}
i.e., the equation of state of RDE will evolve across the
cosmological-constant boundary $w=-1$ (actually, it evolves from the
region with $w>-1$ to that with $w<-1$). That is to say, the choice
of $\alpha<1/2$ makes the RDE today behave as a phantom energy that
leads to a cosmic doomsday (``big rip'') in the future. Thus, as
discussed above, the value of $\alpha$ determines the destiny of the
universe in the holographic RDE model.

According to the observational constraints (from the joint analysis
of data of SN, BAO, and WMAP5) in Ref.~\cite{Zhang:2009un}, the
best-fit result for $\alpha$, with $1\sigma$ uncertainty, is $\alpha
= 0.359_{-0.025}^{+0.024}$, so the RDE takes on the quitom feature.
Now, the question arises. Now that the observations show that the
holographic RDE will more likely behave as a phantom energy, it is
expected that the big rip will occur in a finite time. However, in
the holographic model of dark energy, is the big rip really allowed?
We remember that actually the framework of the holographic model of
dark energy is the effective field theory with a UV/IR duality.
Before the big rip, the UV cutoff of the theory will exceed the
Planck energy scale, and this will definitely spoil the effective
field theory. So, in the holographic model of dark energy, the big
rip is actually not allowed. Now the question is whether we can find
some mechanism to prohibit the occurrence of the big rip in this
holographic model.

In order to avoid the big rip, we appeal to the extra dimension
scenario. We now consider the story of the phantom-like holographic
RDE in a braneworld cosmology. It is anticipated that the extra
dimension effects will become important shortly before the would-be
big rip. We expect that when considering the effects of extra
dimensions, the big rip would be avoided.

Focusing on the case with one extra dimension compactified on a
circle, the effective four-dimensional Friedman equation is
\cite{Cline:1999ts}
\begin{equation}\label{General Fried equ}
3H^2 = \rho \left(1 + \frac{\rho}{\rho_c} \right) \,,
\end{equation}
where $\rho_c=2\sigma$, with $\sigma$ the brane tension,
\begin{equation}
\sigma={6(8\pi)^2M_*^6\over M_{pl}^2},
\end{equation}
where $M_*$ is the true gravity scale of the five-dimensional
theory, and in this expression we explicitly write out $M_{pl}$. In
general, the most natural energy scale of the brane tension is of
the order of the Planck mass, but the problem can be generally
treated for any value of $\sigma>{\rm TeV}^4$.

\begin{figure}[h]
\includegraphics[width=0.4\textwidth]{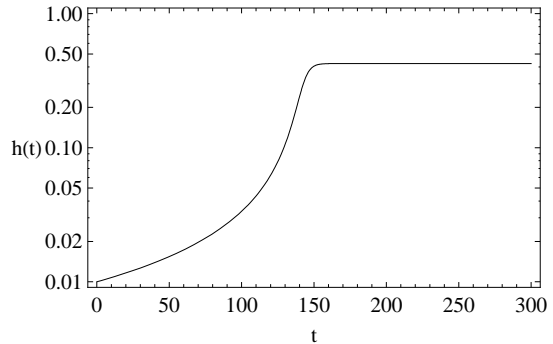}
\caption{\label{fig::ht} Evolution of $h(t)$ with $\alpha = 0.36$.
In this plot, the initial condition is chosen as $h(0) = 0.01$,
where $t=0$ represents the time at which the extra dimension effects
begin to dominate the evolution. Actually, we find that the initial
condition is not important because there is an attractor solution
for the far future evolution, in which the universe will enter the
steady state (de Sitter) finale eventually.}
\end{figure}

\begin{figure}[h]
\includegraphics[width=0.4\textwidth]{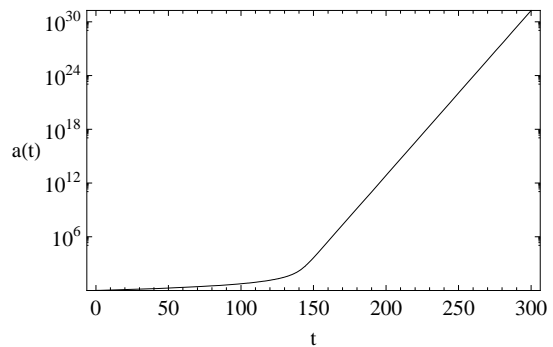}
\caption{\label{fig::at} Evolution of the scale factor $a(t)$. It
should be noticed that the far future evolution is an attractor,
steady state, at which the evolution of the scale factor $a(t)$ is
exponential expansion.}
\end{figure}

When the energy density is larger than $\rho_c$, the second term in
the rhs. of Eq.~(\ref{General Fried equ}) will become significant;
in this situation, taking the energy density of RDE, (\ref{energy
density}), we can analytically solve the equation, and obtain
\begin{equation}\label{steady state analytical}
h(t) = \frac{1}{2\sqrt{3}\alpha +
C_0\exp{\left(-\sqrt{\frac{\rho_c}{3}} \frac{t}{\alpha}\right)}} \,,
\end{equation}
where $h\equiv H/\sqrt{\rho_c}$ and $C_0$ is an integration constant
determined by the initial condition of $h$. Definitely, there is an
attractor solution in the far future. Obviously, the attractor
solution is the steady state (de Sitter) universe. We also solve the
modified Friedmann equation (\ref{General Fried equ}) numerically,
and plot the evolutionary trajectory of $h(t)$ in
Fig.~\ref{fig::ht}. The corresponding plot of $a(t)$ is shown in
Fig.~\ref{fig::at}. The initial condition we choose for $h$ is $h(0)
= 0.01$, in which $t=0$ represents the time at which the second term
on the right-hand side. of the modified Friedmann equation
(\ref{General Fried equ}) becomes important. Although we do not know
what the exact initial condition should be, we can make a choice
$h(0) = 0.01$ to illustrate the future evolution of the universe.
Furthermore, we find that the initial condition is actually not so
important in this case, namely, no matter what the initial condition
is, the universe will enter the steady state eventually. Moreover,
the upper limit value of $h$ even does not depend on $\rho_c$;
actually, it only depends on the parameter $\alpha$,
\begin{equation}\label{upper limit}
h_{max} = \frac{1}{2\sqrt{3}\alpha} \left(1-2\alpha\right)^{1/2} \,
,
\end{equation}
see Eq.~(\ref{steady state analytical}) for a cross check for this
result. For example, when $\alpha = 0.36$, the upper limit for $h$
is $h_{max} \approx 0.42$, and this is confirmed by the numerical
calculation, as shown in Fig.~\ref{fig::ht}.

Although the upper limit of $h$ only depends on the parameter
$\alpha$, the initial condition $h(0)$ does determine the time when
the universe enters the de Sitter space-time, namely, the larger
$h(0)$ is, the later the universe enters the de Sitter space-time.

So far, we only considered the RDE evolution in the far future when
the matter contribution is obviously negligible. However, now, we
try to consider the whole story from today to the far future. In
this case, the matter component should be involved in the
evaluation. When considering the matter, the modified Friedmann
equation (\ref{General Fried equ}) reads
\begin{equation}
3\beta H^2 Y^2 + Y - 1 = 0 \,,
\end{equation}
where $\beta \equiv \rho_c^{-1}$ and $Y \equiv \rho_m/(3H^2) +
\alpha (2+\dot H/H^2)$, for convenience. Solve this equation and
keep only the linear term of $\beta$ for approximation, $Y = 1 - 3
H^2\beta$. Recovering $Y$ and using $\rho_m = \rho_{m0}e^{-3x}$, we
get the differential equation for the Hubble parameter,
\begin{equation}\label{Friedmann approx}
\frac{(r^2)'}{2} + \left(2-\frac{1}{\alpha}\right)r^2 +
\frac{\Omega_{m0}e^{-3x}}{\alpha} + \frac{3r^4}{\alpha}\tilde\beta =
0 \,,
\end{equation}
where $\tilde\beta \equiv H^2_0\beta$. Substituting
\begin{equation}
u(x) = \exp{\left[\frac{6\tilde\beta}{\alpha}\int r^2 dx +
\left(2-\frac{1}{\alpha}\right)x\right]}
\end{equation}
reduces Eq.~(\ref{Friedmann approx}) to a second-order linear
equation of $u$,
\begin{equation} \label{Bessel}
\xi^2 \frac{d^2u}{d\xi^2} + \xi \frac{du}{d\xi} + \left[\xi^2 -
\frac{4}{9}\left(2-\frac{1}{\alpha}\right)^2\right] u = 0 \,,
\end{equation}
where
\begin{equation}
\xi \equiv
-\frac{4}{3\alpha}\left(3\tilde\beta\Omega_{m0}\right)^{1/2}
e^{-3x/2} \,,
\end{equation}
and the solution of Eq.~(\ref{Bessel}) is the linear combination of
Bessel functions,
\begin{equation}\label{sol to Bessel}
u(\xi) = C_1J_{\nu}(\xi) + C_2 Y_{\nu}(\xi),
\end{equation}
where $\nu = 4/3 - 2/(3\alpha)$ and $C_1$, $C_2$ are the integration
constants.

\begin{figure}[h]
\includegraphics[width=0.4\textwidth]{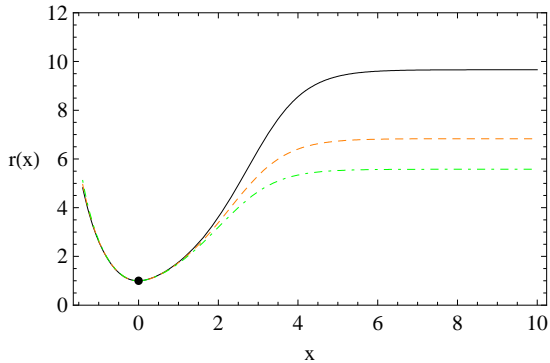}
\caption{\label{fig::fullrx} Evolution of $r(x)$ with $\alpha =
0.36$, and $\tilde\beta = 0.001$ (solid), $\tilde\beta = 0.002$
(dashed), and $\tilde\beta = 0.003$ (dotdashed), respectively. It
should be noted that here the initial condition is $r(0) = 1$, where
$x=0$ ($a=1$) represents today. The black point corresponds to
today's value, $r(0)=1$.}
\end{figure}

As an example, we plot the evolution trajectories of $r(x)$ from the
past to the future, as shown in Fig.~\ref{fig::fullrx}. It should be
noted that here the initial condition is $r(0) = 1$, where $x=0$
($a=1$) represents today.

In addition, we know that the value of $h$ in the far future does
not depend on the value of $\rho_c$, while for $r$ it does. From
Fig.~\ref{fig::fullrx}, one can see that when $\tilde\beta$ is
larger, the future value of $r$ is smaller. In all the cases, the
universe will finally enter into the pure de Sitter space-time and
avoid the big rip.\footnote{For early works on dark energy models
with late-time de Sitter attractor, see Ref.~\cite{XinzhouLi}.}

Furthermore, one may also consider another possible braneworld
scenario, for which the effective four-dimensional Friedmann
equation is \cite{Shtanov:2002mb}\footnote{This modified Friedmann
equation can also arise from the loop quantum cosmology. For
applications in cosmology see, e.g., \cite{Brown:2004cs}.}
\begin{equation}\label{loop}
3H^2 = \rho \left(1 - \frac{\rho}{\rho_c} \right) \,,
\end{equation}
where the negative sign arises from a second timelike dimension.
From the energy conservation law, we obtain
\begin{equation}\label{dot h}
2\dot H = (1+w)\left(\frac{\rho^2}{\rho_c} - 3H^2\right) \,,
\end{equation}
where $w$ is the equation of state parameter for dark energy. Thus,
when $\rho=\rho_c$, we have
\begin{equation}
2\dot H = (1+w)\rho_c \,.
\end{equation}
In usual cases, there is a bounce for a phantom-like ($w<-1$) energy
when $\rho=\rho_c$. We often call this bounce point the
``turnaround'', since at which $H = 0$ and after which $H < 0$, so
the universe turns around from expansion to contraction.

However, in the case of RDE, there does not exist such a turnaround
point. The argument is rather simple. When $H = 0$, the energy
density of RDE is $\rho_{de}=\rho_c=3\alpha \dot{H} > 0$, which
means that when $\rho=\rho_c$, $\dot{H}$ is still positive, so $H$
will still increase and cannot become negative. Therefore, the
universe will never enter into the contracting phase.

Finally, we feel that it would be better to pause for a while to
make some additional comments on the added complication of an extra
dimension. It is well known that the holographic dark energy
provides a natural way out of the fine-tuning and coincidence
problems, as discussed in the literature. However, one might ask
such a question: Whether would these merits be spoiled by the extra
dimension scenario introduced in this Letter? Definitely, the answer
is NO! Actually, the extra dimension introduced here only affects
the evolution of dark energy when the energy density of the universe
is ultra large. There are two such stages in a universe containing
matter and phantom (or quintom): One is the very early beginning of
the universe when dark energy can be neglected, and the other is the
distant future when the universe is dominated by a high-energy
phantom component. Therefore, obviously, the extra dimension does
not affect the way that the holographic dark energy explains the
fine-tuning and coincidence problems at the present epoch of the
universe. Of course, according to the same reason, the extra
dimension does not affect the fitting results of
Ref.~\cite{Zhang:2009un}, either.

Moreover, it should be emphasized that the aim of this Letter is to
explore the possibility of avoiding the theoretical puzzle (namely,
the appearance of the big-rip singularity) in the holographic model
of dark energy (that is based on an effective quantum field theory)
by imposing an extra dimension scenario. The extra dimension effects
play an important role in influencing the future evolution of the
universe when the cosmic energy density becomes ultra high, but do
not impact the past and present universe when the cosmic energy
density is negligible comparing with $\rho_c$. So, the extra
dimension scenario of dark energy discussed in this Letter could not
be distinguished from the one without an extra dimension by using
the current observations.

In conclusion, we have investigated the cosmological evolution of
the holographic Ricci dark energy in the Randall-Sundrum braneworld
scenario. In the holographic Ricci dark energy model, when the
parameter $\alpha<1/2$, the equation of state of dark energy will
cross $-1$ during the evolution, i.e., $w$ will evolve from the
region of $w>-1$ to that of $w<-1$. The observations show that the
parameter $\alpha$ is indeed smaller than $1/2$, so the late-time
behavior of RDE is really like a phantom energy. Therefore, it seems
that the big rip is inevitable in this model. On the other hand,
since the RDE model originates from the holographic model that is
based on the effective field theory with a UV/IR duality, the big
rip seems inconsistent with the theoretical framework. In order to
avoid the big rip, we appeal to the extra dimensions. We consider
the story of the holographic Ricci dark energy in a braneworld
cosmology, and we find that, interestingly, a steady state future
occurs, in stead of the big rip.

\begin{acknowledgments}
We are grateful to Yi-Fu Cai, Chang-Jun Gao, Qing-Guo Huang, Miao
Li, Xin-Zhou Li, Dao-Jun Liu, Yun-Song Piao, Tower Wang, and Yi Wang
for useful discussions. This work was supported by the Natural
Science Foundation of China under Grants Nos.~10705041 and 10975032.
\end{acknowledgments}


\end{document}